\documentclass[12pt]{article}

\pdfoutput=1

\usepackage{a4wide}
\usepackage{amssymb}
\usepackage{amsmath}
\usepackage{bm}

\usepackage{scrextend}
\addtokomafont{labelinglabel}{\sffamily}

\usepackage{hyperref}
 \numberwithin{equation}{section}

\usepackage{amsfonts}

\usepackage{epsfig}

\newcommand{\be}{\begin{equation}}
\newcommand{\ee}{\end{equation}}
\newcommand{\bea}{\begin{eqnarray}}
\newcommand{\eea}{\end{eqnarray}}

\def\tr{{\rm tr}}

\begin{document}

\setcounter{table}{0}

\begin{flushright}\footnotesize

\texttt{ICCUB-19-005}

\end{flushright}

\mbox{}
\vspace{0truecm}
\linespread{1.1}

\vspace{0.5truecm}

\centerline{\large \bf Time-dependent compactification to de Sitter space: a no-go theorem}

\vspace{1.3truecm}

\centerline{
    {\large \bf J. G. Russo${}^{a,b}$}
   {\bf and} 
   {\large \bf P. K. Townsend${}^{c}$}
 }

\vspace{0.8cm}

\begin{labeling}{u}

\item [${}^a$]{{\it Instituci\'o Catalana de Recerca i Estudis Avan\c{c}ats (ICREA),\\
Pg. Lluis Companys, 23, 08010 Barcelona, Spain.}}

\item [${}^b$]{{\it  Departament de F\' \i sica Cu\' antica i Astrof\'\i sica and Institut de Ci\`encies del Cosmos,\\ 
Universitat de Barcelona, Mart\'i Franqu\`es, 1, 08028
Barcelona, Spain. }}

\item [${}^c$]
{{\it Department of Applied
Mathematics and Theoretical Physics,\\ 
Centre for Mathematical
Sciences, University of Cambridge,\\
Wilberforce Road, Cambridge, CB3
0WA, UK.  }} 

\end{labeling}

\noindent {\it E-Mail:}  {\texttt jorge.russo@icrea.cat, p.k.townsend@damtp.cam.ac.uk} 

\vspace{1.2cm}

\centerline{\bf ABSTRACT}
\medskip

It is known that the Einstein gravitational field equations in $D>4$  spacetime dimensions have no
time-independent non-singular compactification solutions to de Sitter space 
if the $D$-dimensional stress tensor satisfies the Strong Energy Condition (SEC).  Here we show, by example, that the 
SEC alone does not exclude time-dependent non-singular compactifications to de Sitter space, in Einstein conformal frame.  However, this possibility is excluded by the combined SEC and 
Null Energy Condition (NEC) because the NEC forces a time-evolution towards a 
singular $D$-metric.\footnote{In the published version of this paper the dominant energy condition (DEC)
was the stated premise but only the weaker NEC was actually used (see ``Note Added" at the end of this paper).}

\noindent

\vskip 1.2cm
\noindent {Keywords: cosmology, compactification, energy conditions}
\newpage


\section{Introduction}
\setcounter{equation}{0}

The observed accelerated expansion of the Universe is consistent with 
a de Sitter (dS) phase, on the assumption of approximate homogeneity and 
isotropy. This fact is in tension with string/M-theory because 
the low energy effective supergravity theories in 10/11 dimensions 
do not admit time-independent compactifications to de Sitter space. 
In most cases this follows from a simple no-go theorem (originally due to Gibbons   \cite{Gibbons:1984kp} and 
rediscovered in a String/M-theory context by Maldacena and Nu\~nez \cite{Maldacena:2000mw}) that 
rules out this possibility for a stress tensor satisfying the Strong Energy Condition (SEC). 

Some types of singularity are innocuous in String Theory and it may be that allowance for them also allows
compactifications to de Sitter space (which we abbreviate to ``dS compactifications'').  Proposals along these lines have been made, 
notably \cite{Kachru:2003aw},  but no consensus on their validity has yet been reached, and there is some recent contrary evidence that 
consigns dS compactifications to a cosmological ``swampland''  \cite{Obied:2018sgi}.

Another way that the Gibbons-Maldacena-Nu\~nez (GMN) no-go theorem 
might be circumvented is to allow for 
time-dependence of the compact space metric. In this case there is an ambiguity
in the metric on the lower-dimensional spacetime, with different choices 
related by a field redefinition involving time-dependent scalar fields. This
ambiguity is resolved by a choice of ``conformal frame'', and the ``Einstein frame'' 
(which results in the absence of any time-dependent function of scalar fields multiplying 
the Ricci scalar in the lower-dimensional Einstein-Hilbert integrand) is the standard 
choice. 

As we emphasised in a previous work \cite{Russo:2018akp}, the Einstein-frame condition involves an integration over the compact space 
but  implementation
of it  has always (to our knowledge) involved a restriction on the integrand that is sufficient but not necessary. 
If the Einstein- frame condition is implemented in this ``unaveraged'' way 
then it is possible to prove that the SEC rules out even time-dependent (non-singular) dS compactifications but this result does not apply
more generally, as we show here by means of an explicit 5D example with a compact space that is topologically a circle. 
The Einstein-frame condition is satisfied in this example, but {\it not}  in  its ``unaveraged'' form. 

This counter-example to the would-be theorem that non-singular dS compactifications are forbidden by the higher-dimensional SEC is, however, non-physical: the 5D stress tensor does not satisfy the Dominant Energy condition (DEC), which is required by causality. This suggests that 
time-dependent non-singular dS compactifications may be ruled out by the SEC and  DEC combined. In fact, our main result is that 
the DEC alone is sufficient for this purpose if the compact space metric is strictly time-dependent, because it then implies a singularity of the higher-dimensional metric.\footnote{As
explained in the ``Note Added'' at end of this paper only weaker NEC is
needed.}

If this new no-go theorem is combined with the GMN no-go theorem then we 
almost have a proof that non-singular dS compactifications are excluded by the SEC and DEC combined. However, the general ansatz for a time-dependent 
dS compactification involves not only a metric on the compact space but also  a  ``warp factor'', and it may be that this is time-dependent even if the compact 
space metric is not.  In this case, which has never previously been considered, we show that the DEC implies an evolution towards a discontinuous warp factor. The corresponding singular 
$D$-metric might be reached only asymptotically,  but we still expect conditions required for the 
low-energy validity of Einstein's  field  equations to be violated at some finite time.


 \section{Warped cosmological compactifications}

It is useful to consider the issues involved in the larger context of  time-dependent compactifications from 
$D$ dimensions to a  general homogeneous and isotropic (FLRW) spacetime of dimension $d<D$, for 
which the standard form of the metric is
\begin{equation}\label{FLRWmetric}
ds^2_{FLRW} \equiv  g_{\mu\nu} dx^\mu dx^\nu  = -dt^2 + S^2(t)\,  \bar g_{ij} dx^i dx^j\, ,  
\end{equation}
where $S(t)$ is the scale factor as a function of FLRW time, and $\bar g_{ij}$ is the metric in local coordinates
$\{x^i; \ i=1,\cdots, d-1\}$ for a maximally-symmetric $(d-1)$-space with constant  curvature $k$.  
There is a time-slicing of the dS universe for which the metric takes this FLRW form for any value
of $k$. For $k=0$, for example, we have
\begin{equation}\label{null}
S= e^{Ht} \qquad (k=0)
\end{equation}
for constant $H$.  The FLRW metric is a solution of the $d$-dimensional Einstein equations for a perfect-fluid source, specified by 
an energy density and pressure.  For a linear equation of state,  the pressure to energy density ratio is a constant $w$, and in this context the SEC (in $d$ dimensions) is equivalent to 
$w\le -1+ 2/(d-1)$ and the DEC (in $d$ dimensions) is equivalent to $|w|\le1$. The $w=-1$ case corresponds to the dS universe.  The $w>-1$ cases correspond to FLRW universes with 
$S\sim t^\eta$, for some constant $\eta$,  and $\eta\le1$ if the ($d$-dimensional) SEC is satisfied. 

Our starting point will be a $D$-dimensional manifold that is  topologically a product of 
this FLRW spacetime with a compact $n$-dimensional manifold $B$ (so $D=d+n$) and, following \cite{Russo:2018akp},  
we consider a general $D$-metric of the form
\begin{equation}\label{Dmetric}
ds^2_D =  \Omega^2(y;t) ds^2_{FLRW} + h_{\alpha\beta}(y;t) dy^\alpha dy^\beta\, , 
\end{equation}
where $h_{\alpha\beta}$ is the metric on $B$ in local coordinates $\{y^\alpha; \alpha=1,\dots,n\}$, and  $\Omega$ is a nowhere-zero `warp factor': a scalar function on $B$.  
The  condition for $ds^2_{FLRW}$  to be an Einstein frame metric for the effective $d$-dimensional gravity theory is \cite{Russo:2018akp}
\begin{equation}\label{Eframe}
\int_B d^ny\,  \sqrt{\det h}\, \Omega^{d-2}  = G_D/G_d\, , 
\end{equation}
where the constant on the right-hand side is the ratio of the Newton constants in the higher and lower spacetime dimensions. 

Taking a time derivative of the Einstein-frame condition we deduce that 
\begin{equation}\label{EF1}
 0=  \int_B d^ny\,  \sqrt{\det h}\, \Omega^{d-2} X \equiv \langle X\rangle\, 
\end{equation}
where 
\begin{equation}\label{X}
X= \frac12 \tr\left(h^{-1}\dot h\right) + (d-2)\left(\dot\Omega/\Omega\right) \, . 
\end{equation}
Following \cite{Russo:2018akp}, we shall refer to $\langle X\rangle =0$ as the ``first-order Einstein-frame condition". Obviously, $X\equiv0$ is not required for $\langle X\rangle =0$, but
it does suffice. 

We are interested in $D$-metrics of the above type that satisfy the Einstein field equations; for an appropriate choice of units these equations are
\begin{equation}\label{Eeqs}
G_{MN} = T_{MN} \qquad \left(M,N=0,1,\dots,D-1\right). 
\end{equation}
We do not need to solve these equations. Instead, we compute the Einstein tensor on the 
left hand side directly from the  metric of (\ref{Dmetric}). We then identify this with the stress 
tensor, which is therefore expressed in terms of the scale factor $S(t)$, the warp factor $\Omega(t,{\bf y})$ and the compact-space metric $h_{\alpha\beta}(t,{\bf y})$.  By construction, this stress tensor supports the given $D$-dimensional spacetime; for a specific $D$-metric it is a 
specific function of time and the coordinates of $B$. One could ask what specific form of matter has this stress tensor,
but an answer to this question is not required for a determination of whether appropriate energy conditions are satisfied. 

We first focus on the SEC. This is a condition on the stress tensor that, given the Einstein equations (\ref{Eeqs}), is equivalent to positivity of the time-time component of the $D$-dimensional Ricci tensor $R_{MN}$.  A computation yields
\begin{eqnarray}
\label{r00}
R_{00} &=&  -(d-1)\left[(\ddot S/S) + (\ddot\Omega/\Omega)  - (\dot\Omega/\Omega)^2 + (\dot\Omega/\Omega)(\dot S/S)\right] -\frac12 \tr\left(h^{-1}\ddot h\right) 
\nonumber \\
&& + \, \frac12(\dot\Omega/\Omega)\tr\left(h^{-1}\dot h\right) + 
 \frac14 \tr \left(h^{-1}\dot h\right)^2  +  (d-1) |\nabla\Omega |^2+\Omega\nabla^2\Omega \, , 
\end{eqnarray}
where $\nabla $ represents the covariant derivative in the internal $B$ space with metric $h_{\alpha\beta}$.  It is implicit here that $\nabla\Omega$ and $\nabla^2\Omega$ are defined. More, generally, we will assume that all components of the $D$-metric are continuous and  at least twice differentiable, with respect to both time and local coordinates for the compact space $B$. 
Also,  as is customarily understood in this context,  we assume that $B$ has no boundary. 

Consider the case of a time-independent metric on $B$; i.e. $\dot h_{\alpha\beta}\equiv 0$. In principle
the warp factor $\Omega^2$ could still be time-dependent, and we shall investigate this possibility later, 
but it is excluded if the first-order Einstein-frame condition (\ref{EF1})  is implemented in an ``unaveraged'' way by setting $X\equiv 0$. It is also excluded by the restriction to ``time-independent 
compactifcations''.  In this case,  $R_{00}\ge 0$ is equivalent to 
\begin{equation}
 d(d-1) \left(\sqrt{\det h} \, \Omega^{d-2}\right) \ddot S/S \le  \sqrt{\det h}\,  \nabla^2 \Omega^d\, , 
\end{equation}
and integration over $B$ yields $\ddot S\le0$, which excludes accelerated expansion, and hence a
dS universe. This is essentially the GMN theorem.

 \section{A time-dependent de Sitter compactification}

We shall now show, by example, that the $D$-dimensional SEC does not forbid {\it time-dependent} non-singular dS compactifications. We start from a 5D metric of the form 
\begin{equation}
    ds^2_5 = \Omega^2(t+y) ds^2_{dS} + \varphi^2(t+y) dy^2\, \qquad \left(y\sim y+2\pi L\right).
\end{equation}
The compact space has the topology of a circle, and $\Omega$ and $\varphi$ are non-zero periodic
functions of $t+y$. A calculation, using the flat-slicing of dS with $k=0$,  shows that 
\begin{eqnarray}
R_{00} &=& -3H^2 -(\ddot \varphi/\varphi) + (\dot\varphi/\varphi)(\dot\Omega/\Omega) - 3(\ddot\Omega/\Omega) + 3(\dot\Omega/\Omega)^2 + H(\dot\Omega/\Omega) 
\nonumber\\
&&+ \ (\Omega/\varphi)^2  \left[(\ddot\Omega/\Omega)- (\dot\varphi/\varphi)(\dot\Omega/\Omega) + (\dot\Omega/\Omega)^2 \right] \, . 
\label{dodtt}
\end{eqnarray}

The Einstein-frame condition is 
\begin{equation}\label{EF2}
    \int_0^{2\pi L} \! dy \, \varphi(t+y) \Omega^2(t+y) = G_5/G_4\, ,  
\end{equation}
but this fixes only the scale of the product function $\varphi\Omega^2$ because
the left hand side is time-independent as a consequence of periodicity.
As explained earlier the Einstein-frame condition implies its ``first-order'' variant $\langle X\rangle=0$, 
even though $X\not\equiv0$ in this example. 

\begin{figure}[h!]
\centering
\includegraphics[width=0.6\textwidth]{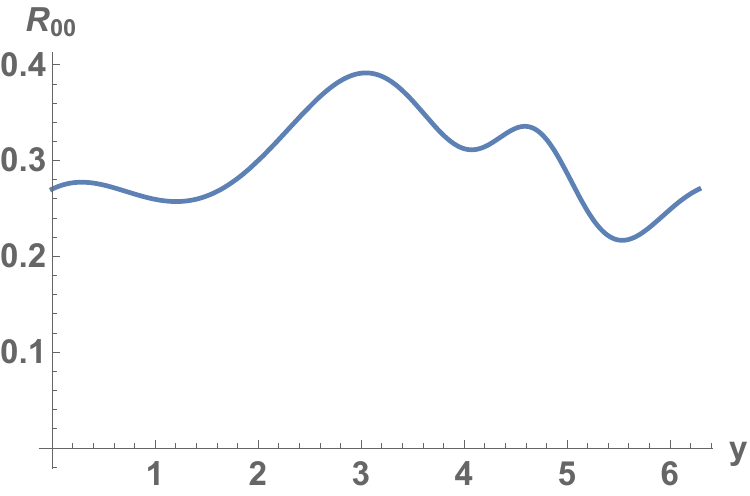}
\caption{  $R_{00}$ of (\ref{dodtt})  as periodic function of $y$ with period $2\pi$;  computed for $L=1$ from (\ref{choice}) with $a=0.2$ and $H = 0.1$.
}
\label{nufa}
\end{figure}

Let us choose
\begin{equation}\label{choice}
    \Omega = 2 A \left\{1+ a\sin\left[ (t+y)/L \right] \right\}\, , \qquad \varphi = A\left\{ 1 + 2a\sin\left[ (t+y)/L\right] \right\}\, , 
\end{equation}
where the  dimensionless constant $A$ is chosen to satisfy (\ref{EF2}). The dimensionless  constant $a$ must satisfy $2|a|<1$ to ensure that $\Omega$ and $\varphi$  have no zeros. For $a=0$ we have $R_{00} =-3H^2<0$,  so the SEC is violated. Otherwise, $R_{00}$ is periodic in $t+y$ and  hence an oscillating function of $y$ for given $t$. The SEC will be satisfied if this function is positive over one period of its argument and it is possible to choose $a$ and $LH$ such that this is the case. An example is $a=1/5$ and $LH=0.1$,  for which $R_{00}$ is plotted in Fig.1,  in units for which $L=1$; we find that 
$R_{00}$ remains  positive for $0<LH<0.24$. Similar examples can be found for toroidal compactification from  any dimension $D>4$. 

To summarise:  the SEC does not forbid time-dependent compactification to de Sitter space.
However, as will be shown below, the 5D stress tensor that supports this SEC-compliant solution 
of the 5D Einstein equations does not satisfy the DEC. 


\section{The DEC  and a no-go theorem}

According to (\ref{Eeqs}), the  $d\times d$ block of the Einstein tensor $G_{MN}$ can be identified with the $T_{\mu\nu}$ components ($\mu,\nu=0, 1, \dots, d-1$) of the $D$-dimensional stress tensor. The result takes the ideal fluid form expected from the FLRW isometries: 
\begin{equation}
    T_{00} =-\rho g_{00} \, , \qquad T_{ij} = P g_{ij}\ , \qquad {i,j=1, \dots d-1}\, . 
\end{equation}
It should be appreciated here that that the density $\rho$ and pressure $P$ are functions on the $D$-dimensional spacetime (in contrast to
the energy density and pressure in the $d$-dimensional FLRW spacetime, which are functions only of time).  The DEC requires $\rho$ to be greater than, or equal to, 
the absolute value of every other component of the ($D$-dimensional) stress tensor. In particular, it requires $\rho \ge |P|$, which 
is equivalent to $\rho\pm P \ge0$ for both choices of sign. These combinations are
 \begin{eqnarray}
\rho -P &=& (d-2)(\ddot S/S) + (d-2)^2(\dot S/S)^2 + (2d-3)(\dot S/S) X +\dot X + X^2 \nonumber \\
&& -\ (d-1) \left[2\Omega\nabla^2\Omega + (d-2)|\nabla\Omega|^2\right] + R(h) + (d-2)^2(k/S^2) \, , \nonumber \\
\rho+P &=& -(d-2)(\ddot S/S)  + (d-2)(\dot S/S)^2 + (\dot S/S)X -\dot X + \frac{1}{(d-2)} X^2 \nonumber \\
&&  - \ \frac{1}{4(d-2)} \left[ \tr\left(h^{-1} \dot h\right)\right]^2 -\frac14 \tr \left(h^{-1}\dot h\right)^2 + (d-2) \frac{k}{S^2}\, , 
  \end{eqnarray}
 where $R(h)$ is the Ricci scalar for the compact space metric, and  $X$ was defined in (\ref{X}). Although $X$ has zero average over $B$ (with weighting by $\sqrt{\det h}\,  \Omega^{d-2}$)
 this is not generally true of $\dot X$ since
 \begin{equation}
 \langle \dot X\rangle = - \langle X^2\rangle \, . 
 \end{equation}

 Notice that $\rho +P$ has no term involving either $\nabla\Omega$ or $R(h)$!  In the dS case, for
 the flat-slicing choice (\ref{null}), the DEC inequality $\rho+P\ge0$ is\footnote{The $k\ne0$ time-slicings of dS, which require a different scale function $S(t)$, yield the same result but with $H$ replaced by
 either $H\tanh{Ht}$ or $H\coth{Ht}$, which has no effect on the arguments to follow.} 
\begin{equation}\label{pkey}
    -\dot X + HX + \frac{1}{(d-2)} X^2 \ge   \frac{1}{4(d-2)}\left[\tr\left( h^{-1}\dot h\right) \right]^2  
    +\frac14 \tr \left[\left(h^{-1}\dot h\right)^2\right] \, . 
\end{equation}
As we are concerned with expanding universes, we may assume here that $H>0$. The left-hand side is zero if $X\equiv 0$, which is a consequence of an 
``unaveraged'' implementation of the Einstein-frame condition; in this case the inequality can only be satisfied if the compact space metric is time-independent. But then we have a time-independent dS compactification, which violates the SEC.

The fact that $\langle X\rangle =0$ tells us that $X$ will take both positive and negative values on $B$ when $X\not\equiv0$. Let us consider the implications of this for
our dS compactification of the previous section, for which $X$ and $\dot X$ are non-zero periodic functions. In one period $X$ will have at least two zeros (exactly two in our example)
 and if  $\dot X$ is negative at one zero then it must be positive at some other zero of $X$. 
It follows that there is a point on $B$ (a circle in our example) for which $X=0$ and $\dot X>0$; but 
the left-hand side of (\ref{pkey}) is then negative, which violates the inequality. The stress tensor implicit in our example therefore violates the DEC. 

In general, at any given time $t_0$, there will be a region of $B$ in which $X>0$, call it $B_+$, and a region in which $X<0$, call it $B_-$.  
There will also be points, and possibly a region, where $X=0$; call this set $B_0$.  This gives us a partition of $B$ into a union of disjoint sets:
\begin{equation}\label{partition}
    B= B_- \cup B_0 \cup B_+\, , 
\end{equation}
where $B_\pm$ are empty sets only when $X\equiv 0$.  This partition will, in general, be time-dependent. 

Let us now suppose that the compact space metric is strictly time-dependent. In this case the DEC  inequality implies the strict inequality 
\begin{equation}\label{key}
    -(d-2)\dot X + (d-2) HX + X^2 > 0\ . 
\end{equation}
Of course, it could happen that the time-dependent compact space metric evolves to one that is time-independent, such that this strict inequality is  replaced by one that allows equality, but we will deal with this possibility below. For now we assume that the strict inequality holds after any sufficiently large time $T$. 
In then follows that $\dot X<0$ on $B_0$ for $t>T$. This implies that all points in $B_0$ at any given time $t=t_0>T$ will be in $B_-$ at $t=t_0+ dt$. 
However, by continuity, we will also have $\dot X<0$ in some neighbourhood of every boundary point of $B_0$. Points in such a neighbourhood that 
are not in $B_0$ will be in $B_-$ or $B_+$, and those in $B_-$ will remain in $B_-$, but  those in $B_+$ that are sufficiently close to the boundary with $B_0$ will move into 
$B_0\cup B_-$.  Since this is true for all $t_0>T$, there will be a flow of points from $B_+$ to  $B_0\cup B_-$, leading to a monotonic decrease in the volume of $B_+$.  Recalling that the average of $X$ on $B$ is zero, we see that a delta-function type singularity of $X$ must form if the volume of $B_+$ shrinks to zero, which is inevitable unless $\dot X \to 0$ on $B_0$, and this can happen only if $\dot h \to 0$ too.

If $\dot h \to 0$ then we have two cases to consider: either the $D$-metric evolves to one for which $X\equiv 0$
(which requires $\dot\Omega \to 0$) or it evolves to one for which $X\not\equiv0$. The former possibility presumes a time-evolution towards a time-independent dS compactification, which is ruled out by the SEC. This conclusion is immediate
if the time-independent dS compactification is reached at finite time, but continuity implies that it is still true if we
suppose that it is reached only asymptotically as $t\to\infty$. 

This leaves only the possibility that the $D$-metric is, or evolves to, one for which the compact space metric is time-independent but the warp factor is still time-dependent, in which case 
$X= (d-2)\dot\Omega/\Omega$. The DEC inequality (\ref{pkey}) is now
\begin{equation}\label{ekey}
 -(d-2)\dot X + (d-2) HX + X^2 \ge 0 \ .  
\end{equation}
On $B_0$ this implies that $\dot X\le 0$. Unless this inequality is saturated, there will again be a monotonic decrease in the volume of $B_+$ but now we may  
suppose, without violating the inequality, that the $D$-metric evolves to one for which $B_+$ has some non-zero volume and   $\dot X=0$ on $B_0$, from which time onwards
the partition (\ref{partition}) is time-independent. 

In the context of this time-independent partition of $B$,  consider points in $B_-$ in the vicinity of $B_0$, where $X=-\epsilon$ for small positive $\epsilon$.  In this case \eqref{ekey} becomes
\begin{equation}
    -\dot X-H\epsilon+ O(\epsilon^2) \geq 0\, , 
\end{equation}
which implies that $\dot X<0$ in this region for $t>t_0$. This implies in turn that the distance between the surface $X=-\epsilon$ and the $B_0/B_-$ boundary, where $X=0$, must shrink to zero, at least asymptotically, leading to a discontinuity of $X$. This discontinuity must either be reached in finite time, implying a singular $D$-metric, or 
asymptotically as $t\to\infty$. In the latter case the $D$-metric itself is not singular, but its singular limit
implies a breakdown of the approximation implicit in our use of the $D$-dimensional Einstein field equations to discuss cosmological compactifications.

\section{Summary and discussion}

Gravity is an attractive force only for matter satisfying the SEC, but this is not a fundamental physical requirement. 
A more important condition physically, for various reasons, is the DEC. The dark energy generally presumed to 
be the cause of the observed dS-like accelerated expansion of the universe is the form of ``matter'' that maximally violates the SEC while 
still satisfying the DEC.  However, effective supergravity theories for String/M-Theory in dimensions $D=10,11$
have stress tensors that satisfy the SEC, and this presents obstacles to the idea that a 4-dimensional dS universe 
could arise from compactification. One such obstacle is the GMN no-go theorem for  time-independent dS compactifications. 

One might suppose,  for any non-singular cosmological compactification, that  the SEC is necessarily satisfied in the lower dimension if it is satisfied
in the higher dimension.  If this were true  it would imply  that compactification to a generic FLRW spacetime is possible only if its expansion is non-accelerating. However, transient acceleration is typical, for reasons  reviewed in \cite{Townsend:2003qv}, where it was conjectured that late-time acceleration might still be excluded. 
This conjecture was proved by Teo for vacuum solutions\footnote{This restriction was not stated as a premise but is essential to the 
proof, as is 
an ``unaveraged'' implementation of the Einstein-frame condition.} of the higher-dimensional Einstein  equations \cite{Teo:2004hq}, but we recently exhibited 
counterexamples for a non-zero stress tensor satisfying both the SEC and the DEC  in the higher 
dimension \cite{Russo:2018akp}. Specifically, we considered compactifications to  power-law FLRW universes,  for which the expansion is driven by perfect fluid matter 
with a linear equation of state, and we showed that the higher-dimensional SEC imposes a lower bound on the pressure to energy density ratio $w$;  for a 4-dimensional universe
this bound is $w>-1/2$, so the expansion is accelerating for $w\in (-1/2,-1/3)$. 

The SEC bound $w>-1/2$ was derived in \cite{Russo:2018akp} in the context of generic time-dependent compactifications to FLRW universes in Einstein conformal frame, 
but with this Einstein-frame condition imposed in the ``unaveraged'' way that has hitherto been standard. However, this implementation of the Einstein-frame condition is
unnecessarily restrictive, and we have  explored here some additional  possibilities that arise when this restriction is relaxed.  An obvious question is whether there 
are additional possibilities with $w<-1/2$. We have not attempted to fully answer this question but we have shown that even dS compactifications, for which $w=-1$, are 
not excluded by the SEC, although the DEC was violated in our example. 

Thus, the SEC is a remarkably weak condition on cosmological compactifications once time-dependence is allowed.  Equally remarkable is the power of 
the DEC in this context: even though it is never violated in the lower dimension, it  forces the higher-dimensional metric to evolve towards a singularity. 
If we demand non-singularity then the combined SEC and DEC in the higher-dimension prevent compactification to dS irrespective of whether this 
compactification is time-independent or time dependent!  This result cannot be derived from either the SEC or the DEC  alone:
we have exhibited  a dS compactification that is compatible with the SEC but  violates the DEC, and if we insist only on the DEC then it is easy to find SEC-violating 
matter that will allow dS-compactification (an example is a positive cosmological constant, which allows a toroidal dS compactification).  

The modern understanding of Einstein's gravitational field equations  is that they arise as an effective description, valid at sufficiently low energy, of some quantum gravity 
theory, possibly String/M-theory. In this context, singularities indicate a breakdown of this low-energy effective description. The GMN no-go theorem 
is therefore a statement about limitations of the effective theory on the assumption that all matter satisfies the SEC: no non-singular time-independent dS compactification metric 
will solve the Einstein field equations.  The no-go theorem proved here, based  on the DEC, makes a subtly different statement: any non-singular time-dependent dS 
compactification metric that solves the  Einstein field equations within some initial time period will have a singularity in its future. 

The broad-brush implication of the DEC for time-dependent dS compactifications is not that it is impossible but that it can only be properly considered in the context of some ultra-violet completion of higher-dimensional General Relativity, such as String/M-theory.

\section*{Note added: DEC vs NEC}

In Section 4 we considered the implications of the DEC for a stress-energy tensor $T_{MN}$ in a $D$-dimensional spacetime with coordinates $x^M=\{t,x^i, y^\alpha\}$ where $i=\{1, \dots d-1\}$, and $\alpha= \{1, \dots D-d\}$.  Here we do the same for the Null Energy Condition (NEC), which is 
$$
T_{MN}n^M n^N \ge0 \, , \qquad g_{MN} n^M n^M =0 \, . 
$$
In the context of Section 4, the stress-energy tensor $T_{MN}$  has a $d\times d$ block that has the form (4.1) of a perfect fluid stress-energy tensor $T_{\mu\nu}$ in a $d$-dimensional spacetime with FLRW metric (2.1) (although the energy density $\rho$ and pressure $P$ depend on all $D$ spacetime coordinates). We may choose the null D-vector $n$ such that 
$$
n^M\partial_M = \partial_t + n^i \partial_i \, , \qquad g_{ij} n^i n^j =1\, . 
$$
For this choice the NEC condition requires
$$
\rho + P \ge 0\, ,  
$$
which is implied by the DEC but is weaker than it. The proof of the no-go theorem in Section 4 
used only this weaker NEC condition (in addition to the SEC). The energy conditions needed for 
the no-go theorem are therefore the SEC and NEC.


\section*{Acknowledgments}

JGR acknowledges financial support from projects 2017-SGR-929, MINECO
grant FPA2016-76005-C.
PKT is is partially supported by the STFC consolidated grant ST/P000681/1.

\setcounter{section}{0}

\end{document}